\documentclass[aps,prl,twocolumn,amsmath]{revtex4-1}

\usepackage{color,graphicx}
\usepackage{bm}% bold math

\usepackage{amsmath}
\usepackage{amsfonts}
\usepackage{amssymb}

\usepackage[latin1]{inputenc}

\begin{document}

\title{Quantum criticality in a uniaxial organic ferroelectric}

\author{S. E. Rowley$^1$, M. Hadjimichael$^1$, M. N. Ali$^2$, Y. C. Durmaz$^{1,3}$, J. C. Lashley$^4$, R. J. Cava$^2$ and J. F. Scott$^1$\\}
\affiliation{1. Cavendish Laboratory, University of Cambridge, J. J. Thomson Avenue, Cambridge, CB3 0HE, England \\2. Department of Chemistry, Princeton University, Princeton, New Jersey, 08545, U.S.A. \\3.Department of Physics, Fatih University, Buyukcekmece Campus, 34500, Istanbul, Turkey \\4. Los Alamos National Laboratory, Los Alamos, New Mexico, U. S. A.}

\date{6$^{\textnormal{th}}$ October 2014}
\pacs{
}

\begin{abstract}
Tris-sarcosine calcium chloride (TSCC) is a highly uniaxial ferroelectric with a Curie temperature of approximately 130K. By suppressing ferroelectricity with bromine substitution on the chlorine sites, pure single crystals were tuned through a ferroelectric quantum phase transition.  The resulting quantum critical regime was investigated in detail - the first time for a uniaxial ferroelectric and for an organic ferroelectric - and was found to persist up to temperatures of at least 30K to 40K. The nature of long-range dipole interactions in uniaxial materials, which lead to non-analytical terms in the free-energy expansion in the polarization, predict a dielectric susceptibility varying as $1/T^3$ close to the quantum critical point.  Rather than this, we find that the dielectric susceptibility varies as $1/T^2$ as expected and observed in better known multi-axial systems.  We explain this result by identifying the ultra-weak nature of the dipoles in the TSCC family of crystals.  Interestingly we observe a shallow minimum in the inverse dielectric function at low temperatures close to the quantum critical point in paraelectric samples that may be attributed to the coupling of quantum polarization and strain fields.  Finally we present results of the heat capacity and electro-caloric effect and explain how the time dependence of the polarization in ferroelectrics and paraelectrics should be considered when making quantitative estimates of temperature changes induced by applied electric fields.

\end{abstract}

\maketitle

There has been a great deal of interest recently in the field of ferroelectric quantum phase transitions \cite{Rowley14, Kim14}.  A chief reason for this is that the properties of ferroelectrics may be readily tuned with gate voltages and strains, which make quantum ferroelectrics and paraelectrics particularly suited for applications in advanced cryogenic electronics.  Highlighting SrTiO$_3$ as an example, one sees that a pristine optically transparent insulator with a band gap of more than 3eV, and a static dielectric constant greater than 10$^4$, may be tuned through an insulator to metal to superconductor transition and back again with the application of just a few volts \cite{Ueno08, Kawasaki11}.  On the other hand modest strains \cite{Haeni04}, chemical doping \cite{Wu06} or isotope substitution \cite{Itoh99, Rowley14}, can induce ferroelectricity in an otherwise paraelectric ground state.  Ferroelectric quantum critical fluctuations have been observed up to temperatures higher than those often seen in other systems, and over a wide range of tuning parameters. Proximity to a displacive ferroelectric quantum critical point where the transverse optical phonon frequency becomes very small, and the dielectric function can rise to very high values, is believed to be of importance in understanding superconductivity in materials such as chemically doped \cite{Schooley64} or ionic-liquid-gated SrTiO$_3$ \cite{Itoh99} and KTaO$_3$ \cite{Kawasaki11}, and in oxide interface materials \cite{Gariglio09} to name just a few.

The nature of quantum criticality in ferroelectrics is strikingly different from that found in other systems, for example magnetic systems \cite{Rowley10, Rowley14, Sachdev99}.  Quantum criticality arises purely from the atomic vibrations of the lattice and not from electronic or spin degrees of freedom.  The excitations around the critical point are propagating bosonic modes (soft transverse optic phonons), which at the critical point are gapless at the zone centre ($q$ = 0). Precessional dynamics and spin flip processes are absent but crucially the free-energy or starting effective action has both short-range and non-analytical long-range interaction terms due to the dipole-dipole interactions present in ferroelectrics.  The long range dipole term is Coulombic in origin rather than Amperian (relativistic) as in the magnetic case, and is thus orders of magnitude stronger.  In multiaxial systems where the polarization is free to point along the different crystallographic directions, the effects of dipole interactions are relatively benign.  Principally they lead to the well known splitting of the frequencies of transverse and longitudinal polar optical phonon modes with respect to the direction of wave propagation $\textbf{\textit{q}}$ \cite{Strukov98, Lines01}.  In a multi-axial displacive ferroelectric material the transverse optical mode disperses as $\Omega^{2}_{q} = \Delta^2 + v^2q^2$ with $\Delta$ going to zero at the critical temperature $T=T_C$.  The parameter $v$ is the speed of sound of the phonons when the gap $\Delta$ vanishes.  The longitudinal optical modes remain finite at the critical point.  In the self-consistent field model (alternatively known as the self-consistent phonon model), which has been shown to be quantitatively applicable without any adjustable parameters in a number of ferroelectric systems \cite{Rowley14}, the correction to the dielectric susceptibility due to quantum critical fluctuations is as follows as explained more fully in Refs. \cite{Rowley14, Rechester71, Khmelnitskii71, Khmelnitskii73, Roussev03, Das09, Palova09, Tokunaga88}.

\begin{equation}
	\delta \chi^{-1} \sim \int^{q_c}_{0} \frac{q^2 n\left(\Omega_q \right)}{\Omega_q}dq
	\label{integral}
\end{equation}
 
In this equation $n\left(\Omega_q \right)$ is the Bose function evaluated at the transverse optical phonon frequency and $q_c$ is a cut-off wavevector typically taken to be the Brillouin zone boundary.  For brevity we have neglected the zero point part which is relatively temperature independent.  As explained more fully in Ref. \cite{Rowley14}, this may be solved numerically over the full temperature range from the quantum regime to the classical regime.  In the classical regime for a material with or without a finite Curie temperature, and well away from the quantum critical point, the model predicts a Curie-Weiss like susceptibility $\chi^{-1} \sim \left( T-T_C \right)$.  Close to the quantum critical point the equations become independent of the cut-off wavevector and lead to a temperature dependence which may be expressed in closed form as follows

\begin{equation}
	\chi^{-1} = a + \frac{5 \epsilon_0 k^{2}_{B} b }{18 \hbar c v}T^2
	\label{T-squared}
\end{equation}

where $a$, $b$ and $c$ are the parameters of the Ginzburg-Landau free-energy expansion in the polarization $P$ at zero temperature, i.e. $f = \left(a/2 \right) P^2 + \left( b/4 \right) P^4 + \left( c/2 \right) (\nabla P)^2$, $\epsilon_0$ is the permitivity of free space, $k_B$ the Boltzman constant and $\hbar$ the Planck constant.

The situation is quite different for the case of a uniaxial system where due to the crystalline details the polarization is confined to vary along one direction only, say the $z$ direction.  In this situation as well as the splitting of the frequencies between transverse and longitudinal optical phonon frequencies, the dispersion of the transverse phonons is modified as follows \cite{Strukov98, Larkin69} 

\begin{equation}
	\Omega^{2}_{q} = \Delta^2 + v^2q^2 + \lambda^2 \left(\frac{q_z}{q}\right)^2
	\label{uniaxial-dispersion}
\end{equation}

in which

$$
\lambda^2 = \frac{4 \pi Q^2}{\mu V_0},
$$
with $V_0$ the volume of the unit cell, $Q$ the effective charge associated with
the soft mode and $\mu$ the reciprocal mass of atoms involved in
the vibrations of this mode. The non-analytic character of the last
term in the right-hand side of Eq. \ref{uniaxial-dispersion} (it does not vanish as the wavevector goes to zero) originates from the long-range character of dipole-dipole interactions. This is the case for both classical and quantum critical points.  Noting that $(q_z / q)^2 = cos^2 \theta$ where $\theta$ is the angle between the $z$ axis and the direction of wave propagation $\textbf{\textit{q}}$, the integral in Eq. \ref{integral} now becomes \cite{Khmelnitskii71, Khmelnitskii14}

\begin{equation}
	\delta \chi^{-1} \sim \int^{q_c}_{0} \int^{1}_{-1} \frac{q^2 n\left(\Omega_q \right)}{\Omega_q}dq d \left( cos \theta \right)
	\label{integral2}
\end{equation}

Again this may be solved numerically giving the full temperature dependence of the susceptibility for uniaxial systems. Effectively the $cos(\theta)$ integral acts as one additional dimension and by an appropriate change of variables the integral may be solved analytically close to the quantum critical point where $\Delta$ approaches zero.  The result in this case is that the inverse susceptibility varies as the cube of the temperature \cite{Khmelnitskii71, Khmelnitskii14}.

\begin{equation}
	\chi^{-1} \sim  T^3
	\label{T-cubed}
\end{equation}

\begin{figure}
\includegraphics[height=2.6in]{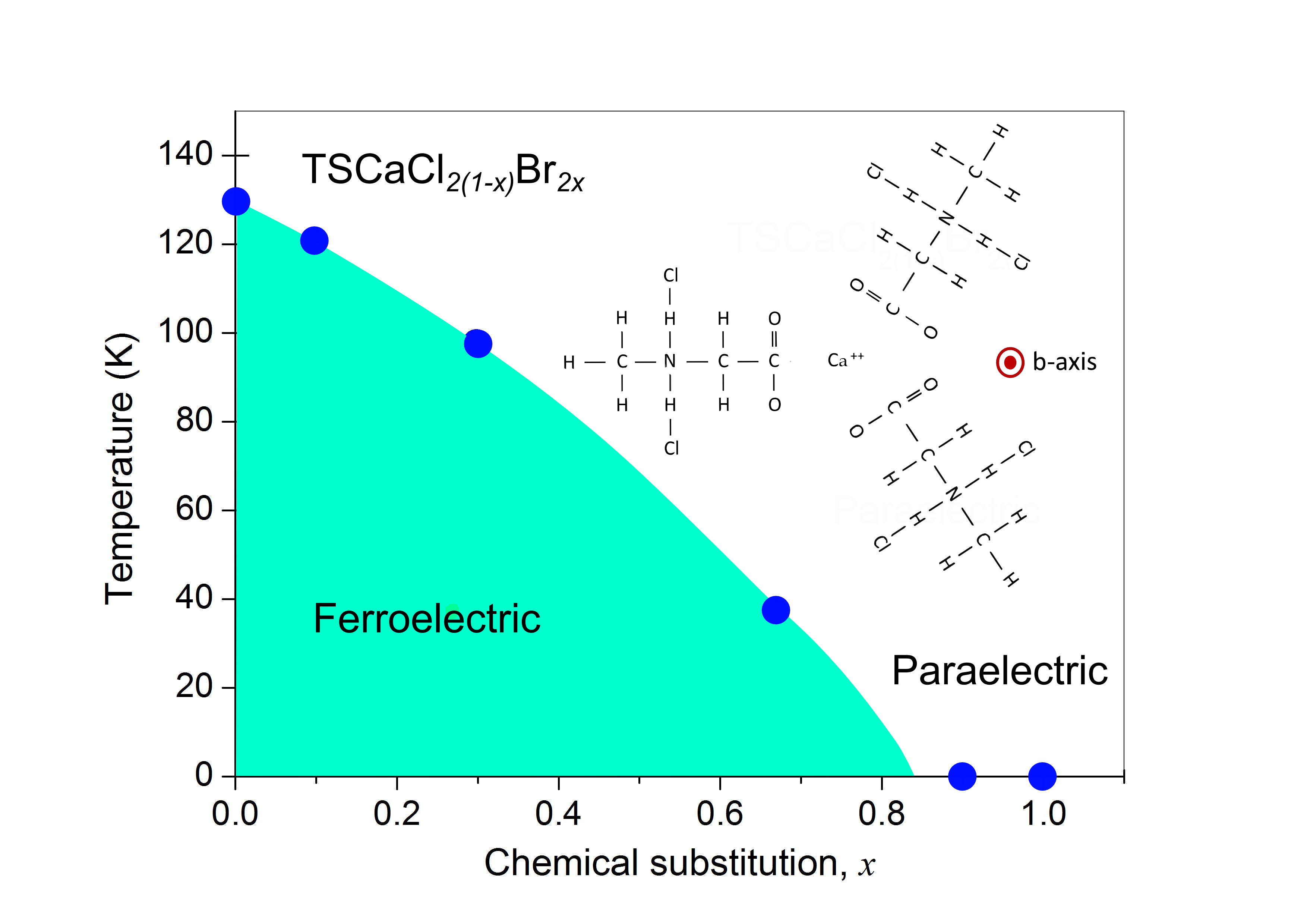}
\caption{\label{phase-diagram} Phase diagram of brominated tris-sarcosine calcium chloride.  The shaded region is ferroelectric and the unshaded region is paraelectric.  The two phases are separated at zero temperature by a quantum critical point between approximately 80 and 90 per cent bromine substitution.  `TS' stands for tris-sarcosine, i.e. (CH$_3$NHCH$_2$COOH)$_3$.  The blue dots show the values of $T_C$ and $x$ in the samples measured in this study.  The upper right inset shows a schematic of a region of the crystal structure of TSCC in the a-c plane.  The static and fluctuating polarization is confined to the $b$ direction, i.e. pointing out of the page, which makes TSCC a uniaxial ferroelectric.}
\end{figure}

Ferroelectricity in TSCC may be tuned to absolute zero using bromine substitution as shown in Fig. \ref{phase-diagram}.  The ferroelectric transition in TSCC is observed to be second order \cite{Ashida72} in all experiments carried out so far, an uncommon feature among ferroelectrics, which implies that the zero temperature transition is a quantum critical point (QCP) with quantum fluctuations persisting over a range of temperatures and doping levels.

TSCC is a uniaxial ferroelectric with the polarization forming along one direction only (the b-axis as indicated in the inset to Fig. \ref{phase-diagram}).  It is one of the most prototypical displacive ferroelectrics known,\cite{Feldkamp81} in the sense that there exists an under-damped soft mode that can be followed into the low GHz frequency regime \cite{Mackeviciute13} from high-$T$ values of $ca.$ 630 GHz = 21 cm$^{-1}$. Deuteration \cite{Gergs86} produces little change in $T_C$, implying that the transition is not controlled by the N-Cl-hydrogen bonds, compatible with the soft-mode mechanism operating in TSCC \cite{Kozlov83}.  The Curie constant of pure TSCC is less than a hundred Kelvin, the smallest known Curie constant of any known ferroelectric. By way of comparison, typical values for oxide perovskites such as BaTiO$_3$ are 50,000 K. This puts TSCC in the family of ultra-weak ferroelectrics. An unrelated form of weak ferroelectricity arises in certain multiferroics where small polarizations are induced by magneto-electric coupling.

\begin{figure}
\includegraphics[height=7.4in]{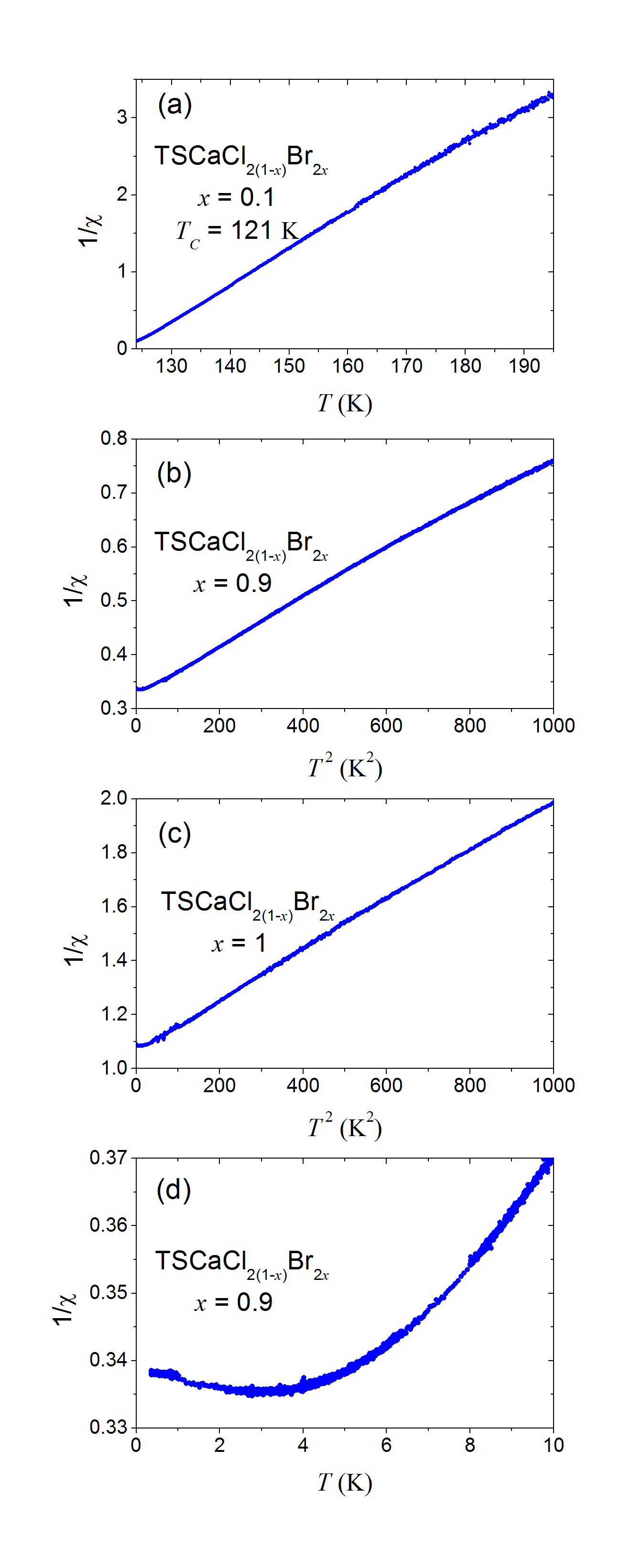}
\caption{\label{dielectric} The inverse dielectric susceptibility is shown for three representative bromine doping levels in TSCC. In (a) $1/ \chi$ is plotted for the 10 per cent Br doped sample as a function of temperature and shows approximately linear, i.e. Curie-Weiss like dependence above $T_C$. In (b) and (c) $1/ \chi$ is plotted against $T^2$ for the 90 and 100 per cent Br doped samples which have paraelectric ground states, are close to the quantum critical point, and show an approximately $T^2$ temperature dependence. In (d) the low temperature region of $1/ \chi$ is magnified for the 90 per cent sample (i.e the sample closest to the quantum critical point) and exhibits a shallow minimum around 3 K which was not observed in any of the other doping levels measured.}
\end{figure}

The results of dielectric susceptibility measurements taken in a range of brominated TSCC single crystals are presented in Fig. \ref{dielectric}.

The figures show classical Curie-Weiss behaviour in (a) for the 10 per cent bromine doped sample.  On the paraelectric side, close to the quantum critical point, $T^2$ behaviour is seen in the 100 per cent sample and over a wider temperature range in the 90 per cent sample which is paraelectric but right on the boundary of the ferroelectric state at zero temperature. The cross-over temperature $T^*$ above which the $T^2$ power law changes to a lower power was taken to be the point where the exponent $\gamma$, ($1/ \chi \sim T^{\gamma}$) deviated from 2 by 10\%. This was approximately 30K in the 90\% Br sample and 25K in the 100\% Br sample.  Interestingly, in the 90\% sample, but absent in the 100\% sample, we observe a shallow minimum in the inverse susceptibility around 3K (Fig. \ref{dielectric}d).  This has been seen in other quantum critical ferroelectrics such as SrTiO$_3$ and KTaO$_3$ \cite{Rowley14} and has been explained in those cases quantitatively without adjustable parameters as arising from the electro-strictive coupling of the polarization to strain.  This comes from the addition of a term $-g \nabla \phi P^2$  in the free-energy where $\nabla \phi$ is the strain and $g$ the electro-strictive coupling constant \cite{Khmelnitskii71, Palova09, Rowley14}.  This gives rise to a further correction to the inverse susceptibility varying as $-T^4$ at low temperatures in leading order in $T$ which leads to a minimum \cite{Khmelnitskii71}.  The theory predicts that the minimum should be present in paraelectric samples close to the quantum critical point and so could likely be the explanation for the observations presented here.

\begin{figure}
\includegraphics[height=2.5in]{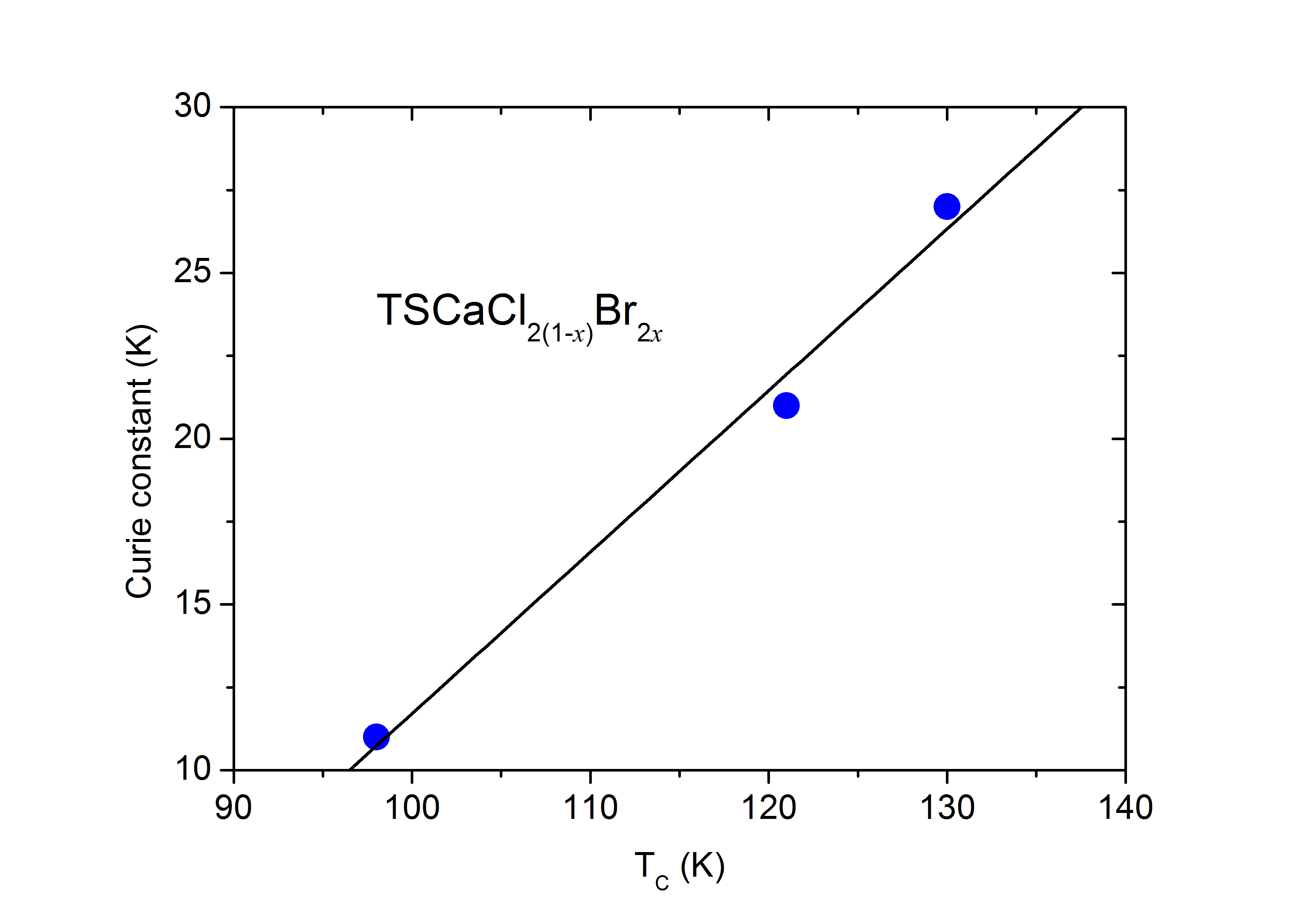}
\caption{\label{curie-constant} The Curie constant versus Curie temperature in samples of TSCC with varying Br content. The Curie constant was determined from a fit to the part of  $1/\chi$ which varied linearly with $T$, i.e. in the classical Curie-Weiss regime of the inverse susceptibility. This held all the way to down to $T_C$ to within one or two degrees for the 0\%, 10\% and 30\% samples and down to the quantum regime, designated by a cross-over temperature $T^*$, in the remaining samples.  The error in determining $C$, which dropped as $T_C$ was supressed, became much larger for the samples with paraelectric ground states due to the tiny magnitude of the dipoles present in them, a characteristic of the TSCC family of materials.  The trend observed in the measured data here was also confirmed from published measurments (e.g. Ref. \cite{Fujimoto81}) of the spontaneous polarization $P_S$ for different bromine doping levels using the relation $C(T_{C,1}) / C(T_{C,2}) = P^2_S(T_{C,1}) / P^2_S(T_{C,2})$ from Eq. \ref{C-bare}.}
\end{figure}

Since TSCC is a uniaxial system one might have expected the inverse susceptibility to vary as $T^3$ close to the quantum critical point as explained above.  A $T^3$ inverse susceptibility (i.e. $\gamma = 3$) has thus far not been observed in nature near to a ferroelectric quantum critical point.  Instead of this we observe a $T^2$ dependence as expected and observed in several multi-axial ferroelectrics close to quantum criticality. We can understand this result by noting that the magnitude of the dipoles in TSCC are extremely small and thus the effects of the dipole interaction term is greatly attenuated.  This is evidenced by a number of observations in TSCC. Firstly, the splitting of the frequencies of transverse (TO) and longitudinal (LO) modes in TSCC is very small ($\sim 50GHz$) and as the TO mode approaches zero at the Curie point the LO mode also drops and traces the temperature dependence of the TO mode (but remains finite) \cite{Prokhorova80, Feldkamp80, Kozlov83}.  The size of the spontaneous polarization in all the ferroelectric samples \cite{Fujimoto81} , and the peak value of the dielectric constant in all the samples measured, is very small (less than 10).  Moreover the peak value gets progressively smaller as the level of bromine doping increases and as $T_C$ is progressively suppressed towards zero \cite{Lashley14}.  As mentioned in the introduction, the Curie constant ($C$) in TSCC is less than 100K and is the smallest of all the known ferroelectrics highlighting the fact that the dipoles are very small in magnitude.  In fact along with the peak value of the dielectric constant, the Curie constant gets progressively smaller (approximately linearly), as a function of Br doping.  This means that the magnitude of the dipoles which started off as being ultra-weak in pure TSCC, get progressively weaker still as one approaches the quantum critical point with Br substitution.  This is illustrated in the plot of Curie constant versus Curie temperature in Fig. \ref{curie-constant}.

All this essentially means that the $\lambda^2(q^2_z/q^2)$ term in Eq. \ref{uniaxial-dispersion} is much smaller than $v^2q^2$ term over the range of wavevectors integrated over in Eq. \ref{integral2}. This means that the correction to the susceptibility given by Eq. \ref{integral2} reduces to the integral Eq. \ref{integral} resulting in an inverse susceptibility varying as $T^2$ as observed, rather than a $T^3$ dependence as expected more generally for uniaxial systems.

The following analysis outlines a simplified model in which $C$ (and thus $p$ and $\lambda^2$) get smaller as $T_C$ is suppressed.

In comparison with ferromagnets, where the Curie constant in the Weiss theory is not an independent parameter but is given by

\begin{equation}
	C = Ng^2 \mu_B^2 S(S+1)/(3k_B)
	\label{C-mag}
\end{equation}

where $g$, $\mu_B$, and $k_B$ are the gyromagnetic ratio, Bohr magneton, and Boltzmann constant, in ferroelectrics $C$ is usually determined empirically by the relation

\begin{equation}
	\epsilon = \epsilon_0 + \frac{C}{T-T_C}. 
	\label{Curie-Weiss}
\end{equation}

Some authors \cite{Tokunaga87, Tokunaga88, Fujimoto81} have defined a `bare' Curie constant

\begin{equation}
	C_0 = N p^2 / k_B, 
	\label{C-bare}
\end{equation}

where $p$ is the dipole moment and $N$ the number of dipoles and term the ratio $C/C_0$ the Rhodes-Wohlfarth ratio, in analogy with ferromagnets.  

In order to describe Rochelle Salt, P. Kobeko and I. Kurchatov \cite{Kubeko30} assumed a rotating rigid dipole model in which the polarizability 

\begin{equation}
	\alpha = p^2/(3 k_B T),
	\label{alpha}
\end{equation}

where each rotating dipole has dipole moment $p$.  The internal field is given by $E_i = E(applied) + \beta P$ where $\beta$ is the relevant Lorentz factor and $P$ the polarization (dipole moment per unit volume).  This leads to the result $C = \sigma T_C$ where the constant parameter $\sigma$ depends on $\beta$ and the system of units in use.  The data for $C(T_C)$ in TSCC/TSCB are shown in Fig. \ref{curie-constant}.  Hence in TSCC or its Br-isomorphs, $C$ decreases towards zero as bromination increases to the QCP.  This means that the magnitude of the dipole moments are suppressed as one approaches the QCP and the assumption of a finite $\lambda^2$ is not possible, meaning that brominated TSCC may behave as an isotropic material asymptotically as $T_C$ approaches zero \cite{Scott89}.

\begin{figure}
\includegraphics[height=2.7in]{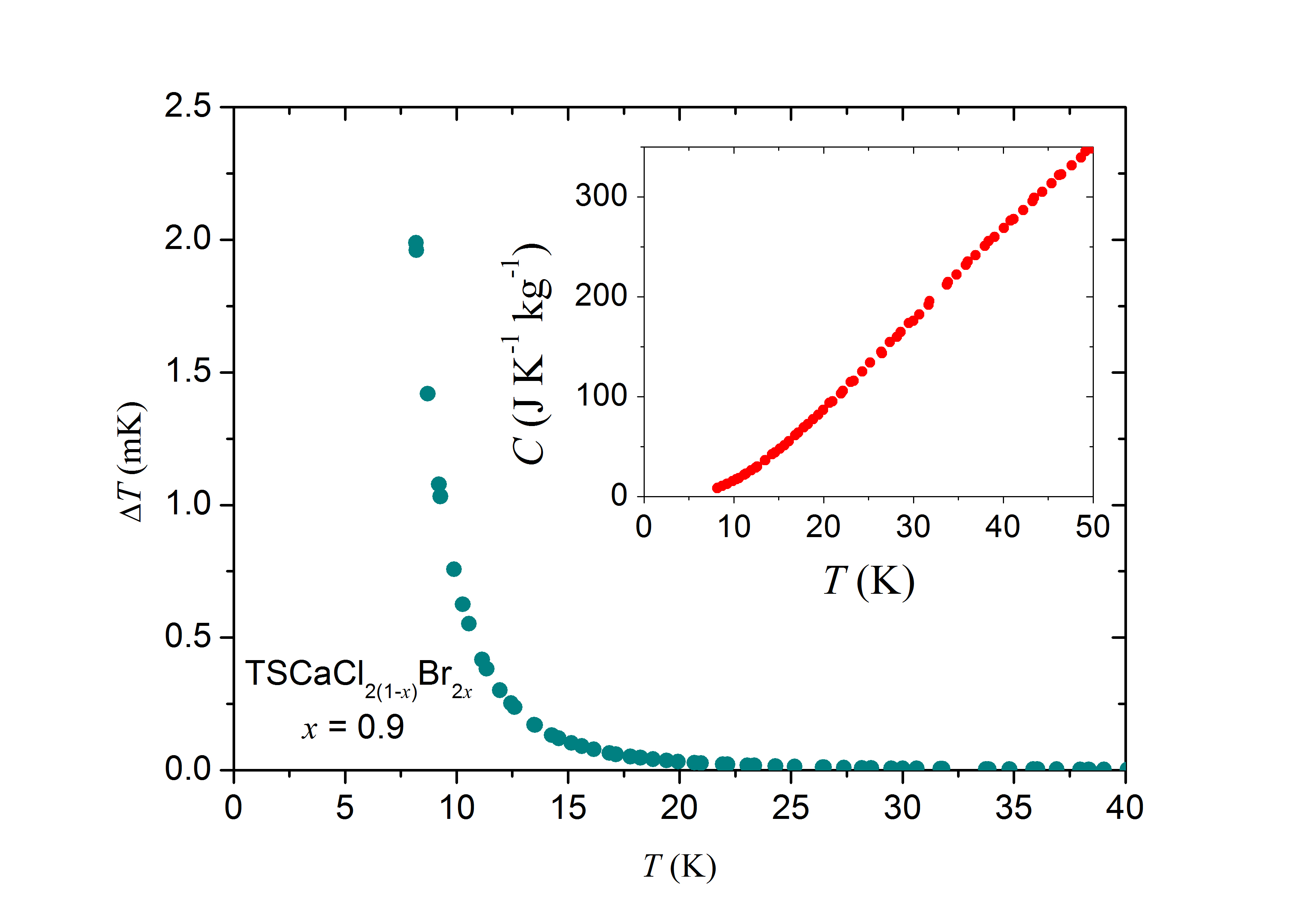}
\caption{\label{electro-caloric} The main figure shows the electro-caloric temperature change achievable at different temperatures for fields of approximately 3 kV/cm calculated using Eq. \ref{temp-change} in the main text for the sample closest to the quantum critical point, i.e. the 90\% Br doped sample of TSCC. The temperature change is enhanced close to the QCP where the dielectric constant is largest and the heat capacity is small but overall the cooling temperatures possible are small due to the ultra-weak nature of the dipoles and the resulting insensitivity of the entropy to applied electric fields.  The measured heat capacity used to estimate the electro-caloric effect is shown in the inset.}
\end{figure}

We now turn to a comment about the electrocaloric effect which is sometimes predicted to diverge as $T$ approaches zero (especially close to a quantum critical point where the dielectric function may be large at zero temperature).  This prediction is due to the fact that from the Maxwell relations, in equilibrium, there is an indirect way of determining the temperature cooling $\Delta T$ \cite{Correia14, Lu2010a, Lu2010b, Mischenko06}:

\begin{equation}
	\Delta T = -T \int \frac{1}{C} \frac{dP(T,t)}{dT}  dE 
	\label{temp-change}
\end{equation}

and this expression contains a specific heat term (for which the results of our measurements in TSCC are shown in the inset to Fig. \ref{electro-caloric}) that diverges faster than its linear pre-factor of $T$. However that assumes that the dielectric function and corresponding polarization $P$ are large.  In the present case they are exceptionally small, and the numerical values of the electro-caloric effect are extremely small, even at low $T$ (Fig. \ref{electro-caloric}).

Also of interest is the time dependence of these relations.  The estimate in Eq. \ref{temp-change} is from the Maxwell relations and generally agrees within a factor of 2 or better with direct measurements of $\Delta T$ \cite{Correia11, Kutnjak14}.  However, it is not exact.  It is based upon equilibrium thermodynamics (dubious in the case of relaxors), and it ignores time dependences.  $P(T)$ actually depends upon time for real ferroelectrics due to relaxation processes \cite{Dietz95}.  After a voltage is applied $P(T,t)$ decays with time $t$, on several different time scales, depending upon temperature $T$.  Typically the long-term value of polarization $P$ is ca. 50 per cent of the short-term value.  In some modern test equipment, the remnant polarization $P_r$ at long times is denoted with a circumflex.

Thus if $\Delta T$ is measured on a time scale long compared with the relaxation time $\tau$, care must be taken to use the proper value of $P$ in Eq. \ref{temp-change}, which may be only half that of the short-time measured electrical value from the $I(V)$ hysteresis curve.  This will generally lead to an overestimate of $\Delta T$ inferred from the `indirect' method described by Eq. \ref{temp-change}, and by as much as 50 per cent.  No published electrocaloric data include this time dependence of the Maxwell relations.  To be more self-consistent, $\Delta T$ must be measured as a function of time $t$ in applying Eq. \ref{temp-change}, and the time scale must be commensurate with the relaxation time $P(T,t)$.

In conclusion we have made detailed measurements and analysis of quantum criticality in an organic and a uniaxial ferroelectric.  At first sight one may have expected the long-range dipole interactions, which are of crucial importance in uniaxial systems, to have drastically changed the power law behaviour of the temperature dependence of the inverse dielectric constant.  This is because for such systems, in general, the effective dimension $d_{eff} = 5$ rather than the usual $d_{eff} = d + z = 4$ for quantum multiaxial systems, where $d=3$ is the spatial dimension and $z=1$ is the dynamical exponent \cite{Rowley14}.  Instead we found in the special case of the TSCC family of materials that due to the ultra-weak nature of the dipoles, and their further attenuation on approaching the quantum critical point, that the system behaves as an isotropic/multiaxial material especially as $T_C$ approaches zero.  The search for $\chi^{-1} \sim T^3$ in a uniaxial quantum critical ferroelectric continues, which has so far remained elusive.  The upturn observed in the inverse dielectric constant of paraelectric samples on the border of ferroelectricity at low temperature may be attributed to the electro-strictive coupling of quantum polarization and strain fields as recently discovered in other systems \cite{Rowley14}.  We found the estimated electro-caloric temperature change to be small (of the order of a few milli-Kelvin over the accessible temperature and field ranges) close to the quantum critical point due to the fact that TSCC is a system with extremely small dipoles.  For these reasons the entropy change on the application of external electric fields is small.

\section{Acknowledgments}
The authors wish to express their thanks for useful help and discussions to P. Chandra, D. E. Khmelnitskii, G. G. Lonzarich and S. S. Saxena.

\section{References}

\bibliographystyle{apsrev4-1}
\bibliography{TSCCPRL2014}

\end{document}